# Never mind the Pollocks

**We investigate the hypotheses that Jackson Pollock's drip paintings are fractals produced by the artist's Lévy distributed motion and that fractal analysis may be used to authenticate works of uncertain provenance[1-5]. We find that the paintings exhibit fractal characteristics over too small a range to be usefully considered fractal; their limited fractal characteristics are easily generated without Lévy motion, both by freehand drawing and Gaussian random motion. Several problems must be addressed to assess the validity of fractal analysis for authentication.**

An image is considered fractal if, when covered with a grid of square boxes of size $L$[1,6] , the number of filled boxes $N \propto L^D$ . Here $D$ , the fractal dimension, is non-integer. It is generally accepted, e.g. by both sides of a recent debate[7,8], that to establish power-law behaviour the range of box sizes must span more than one or two orders of magnitude. For a Pollock drip painting, the range is typically limited to three orders of magnitude by the ratio of the canvas size to the smallest feature size. For all Pollock paintings examined so far, the box-counting curve, a plot of $N$ vs $L$, is found to be a broken power-law[5] ( $N \propto L^{D_L}$ for $L > l_T$, and $N \propto L^{D_D}$ for $L < l_T$ where $l_T$ is a characteristic length). Thus there are two power-laws and less than two orders of magnitude to establish each. Some pitfalls of inferring fractal behaviour from such a limited range are demonstrated below.

For multi-coloured, multi-layered drip paintings (a large part of Pollock's oeuvre) it's claimed[4] that the visible part of each layer separately, and the composite, are all fractals. This claim is presumably an artefact of the limited range of data since it can be shown to be mathematically impossible by consideration of a model in which the individual layers are ideal fractals[6], e.g., Cantor dusts or Sierpinski carpets. Explicit calculation then shows that the box-counting curves for the visible parts of each layer and the composite painting are not power laws even though the individual layers are perfect fractals (see Figure 1).

Taylor *et al.* assert that the "defining visual character of Pollock's drip-paintings is their fractal nature"[4] and "fractals arise from the specific pouring technique developed by Pollock"[9]. To test this claim we drew a number of freehand sketches. Figure 2 shows one we dubbed *Untitled 5,* and its box-counting curve. By criteria espoused in Refs. (2-5), it is a high quality fractal. Fig. 2 undermines the claim[2-5,9] that Pollock's works derive their artistic merit from their (limited) fractal content.

There are insufficient data[10,11] to directly determine whether Pollock's motion whilst painting constituted a Lévy flight. Lévy motion has been inferred[2-5] from the known result that Lévy flights leave a fractal trail[6]. However, the box-counting data of a simulated Gaussian random walk of 100 steps, examined over only one and a half orders of magnitude, are well-described as a fractal ($D = 1.35$ in our simulation; see Fig. 1)— although sufficiently long Gaussian random walks are rigorously known to be non-fractal with dimension $D = 2$. Thus the limited fractal content of Pollock's work does not require

Lévy flights for its explanation. Furthermore Lévy motion has no natural length scales; Pollock's paintings and motion appear to have many[4].

Box-counting authentication assumes that the parameters $l_T$, $D_D$ and $D_L$, and $\chi^2$ (see Figure 2 caption) show characteristic trends that distinguish Pollock from imitators[5,9]. For multifractals[12] and non-fractals $\chi^2$ is not an intrinsic characteristic of the image, but is $C$–dependent (see figure caption) . If Pollock's paintings are multifractal, as claimed in Ref (13), it would be imprudent to use $\chi^2$ as a characteristic parameter. Also, we question the presumption that parameters like $l_T$ for Pollock's works are essentially random variables, circumscribed by a range determined[9] by examination of just 17 drip paintings (out of approximately 180). Further study of systematic effects is needed. Since the values measured by Taylor *et al.*[9] are being kept confidential, a consensus on the limiting range, if there is one, can emerge only after other groups have replicated and extended the box-counting dataset. Finally we note that *Untitled 5* fulfils all criteria used in box-counting authentication that have been made public.

Katherine Jones-Smith & Harsh Mathur

Department of Physics, Case Western Reserve University, Cleveland, Ohio 44106

Correspondence and requests for materials should be addressed to Harsh Mathur. email: hxm7@case.edu

1. Abbott, A., Fractals and art: In the hands of a master, *Nature* **439**, 648-650 (2006).

2. Taylor, R.P., Micolich, A.P., & Jonas, D., Fractal Analysis of Pollock's Drip Paintings, *Nature* **399**, 422-422 (1999).

3. Taylor, R.P., Micolich, A.P., & Jonas, D., Fractal Expressionism: Art, Science and Chaos, *Physics World* **12**, 25-28 (1999).

4. Taylor, R.P., Micolich, A.P., and Jonas, D., The Construction of Pollock's Fractal Drip Paintings, *Leonardo* **35**, 203-207 (2002).

5. Taylor, R.P., Fractal Expressionism—Where Art Meets Science, pp 117-144 in: Casti, J., and Karlqvist, A. (Eds.), *Art and Complexity* (Elsevier Press, Amsterdam, 2003).

6. Mandelbrot, B.B. *The Fractal Geometry of Nature*. W.H. Freeman Co., New York (1977).

7. Avnir, D., Biham, O., Lidar, D. & Malcai, O., Is the Geometry of Nature Fractal? *Science* **279**, 39-40 (1998)

8. Mandelbrot, B.B., Is Nature Fractal? *Science* **279**, 783-783 (1998).

9. Taylor, R.P. *et al*., Authenticating Pollock Paintings Using Fractal Geometry, to be published in *Pattern Recognition Letters*. (Available at http://materialscience.uoregon.edu/taylor/art/TaylorPRL.pdf)

10. Namuth, H. and P. Falkenberg, *Jackson Pollock* (film), Collectors Series, The Museum of Modern Art, New York, N.Y. 1950.

11. Landau, E.G. *Jackson Pollock.* Thames and Hudson, London, 1989.

12. Ott, E. *Chaos in Dynamical Systems*. Cambridge University Press, Cambridge (2002).

13. Mureika, J.R., Cupchik, G.C. and Dyer, C.C., Multifractal Fingerprints in the Visual

**Figure1. Fractal barcode and Gaussian walk.**  (a) A middle-third Cantor dust anchor layer (blue) with a second Cantor dust overlaid (red). Half-blue/half-red bars correspond to the intersection of the dusts; purple, to the union. The upper graph shows box-counting curves: blue dust (shown in blue), red dust (red), the uncovered part of the blue dust (green) and the composite (purple). The curvature of the traces shows that the uncovered portion of the blue layer and the composite are not true fractals. To highlight the curvature of the composite, the lower graph shows the difference of the (rigorously linear) blue and purple curves.  (b) A linear fit to the box counting curve of a hundred step Gaussian walk has a slope of 1.35 with standard deviation $\chi$= 0.025.

*Methods*: (a) The blue dust is obtained by repeatedly dividing segments into three parts and retaining only the first and third; the red by dividing into nine parts and retaining the first, fifth and ninth parts. The 'fractal barcode' shows the appearance of the dusts after four iterations.  (b) Step size = 0.09 x frame width. Smallest box size = 3 pixels. Sizes range over 1.4 orders of magnitude with magnification C = 1.12 (see Fig. 2 for definition).

**Figure 2. *Untitled 5* and its box-counting curve.** The best broken power law fit to these data correspond to slopes of $D_D = 1.53$ and $D_L = 1.84$. The break occurs at $\ln L \approx 4$. The standard deviation of the data from the fit, $\chi$= 0.022. Thus *Untitled 5* fulfils all the criteria used in box counting authentication that have been made public: it has a broken power law behaviour with $D_D$ <$D_L$ and, for a magnification factor *C* (defined under methods) similar to that used by Taylor *et al.,* a $\chi^2$ value in the "permissible range" 0.009 < $\chi$ < 0.025. *Methods:* All our sketches, including *Untitled 5,* are freehand drawings made in Adobe Photoshop using a 14pt Adobe Photoshop 'paintbrush'. The paintbrush leaves a mark when dragged continuously across the 'canvas' via computer mouse. Although not drip paintings, these drawings are human generated patterns, not computer generated, in the sense that these terms are commonly understood and are used by Taylor *et al.*[4] The box sizes were chosen to be $hC^n$ with $n = 1,2,3,....,M$. Here $h$ = 3 pixels is the smallest box size, the magnification factor $C = 1.12$, the cutoff M determines the biggest box size, and the box sizes range over approximately 1.9 orders of magnitude.